\documentclass[
  aps,
  pra,
  twocolumn,
  longbibliography
]{revtex4-2}

\usepackage{amsmath,amssymb,bm,graphicx,hyperref}
\usepackage{booktabs,braket,mathtools,mathrsfs}

\begin{document}

\title{Resonant two-cluster scattering in a quasi-one-dimensional Bose gas}

\author{Tomohiro Tanaka}
\author{Yusuke Nishida}
\affiliation{Department of Physics, Institute of Science Tokyo, Ookayama, Meguro, Tokyo 152-8551, Japan}

\date{March 2026}

\begin{abstract}
    We investigate two-cluster scattering in a quasi-one-dimensional Bose gas.
    We focus on the effective three-body interaction induced by transverse confinement, 
    which is the leading term for breaking integrability in the quasi-one-dimensional setting.
    Exploiting the L\"uscher formula and the integrability of the Lieb-Liniger Bose gas,
    we find a finite and positive scattering length for elastic two-cluster scattering.
    The resulting scattering lengths indicate the emergence of a resonance.
\end{abstract}

\maketitle

\section{Introduction}
One-dimensional systems have been investigated intensively from a theoretical perspective~\cite{RevModPhys.83.1405,RevModPhys.85.1633,KorepinBOOK,Takahashi1999}. 
The Lieb-Liniger model is a paradigmatic model describing a one-dimensional Bose gas with a two-body contact interaction~\cite{lieb:1963,lieb:1963-2,KorepinBOOK,Takahashi1999}. 
This model is exactly solvable by the Bethe ansatz, through which its wave functions and energy eigenvalues can be obtained exactly~\cite{Bethe1931,lieb:1963,lieb:1963-2,KorepinBOOK,Takahashi1999,Gaudin2014BetheWavefunction}.
In addition, the matrix elements of local operators admit concise analytical expressions\cite{KorepinBOOK,Piroli_2015,Slavnov1990_TMP82_273,Kojima:1997detcorr,Pozsgay_2011}.
In the attractive regime,
the Lieb-Liniger model supports bound clusters,
which are referred to as strings~\cite{10.1063/1.1704156,Takahashi1999,Calabrese_2007,PhysRevLett.98.150403}.
Due to the integrability of this model,
scattering between strings is elastic and reflectionless,
and cluster breakup or recombination does not occur~\cite{PhysRev.168.1920}.

One-dimensional systems are not only of theoretical interest but have also been realized experimentally with ultracold atoms~\cite{RevModPhys.83.1405,PhysRevLett.87.130402,PhysRevLett.91.250402,Paredes2004TonksGirardeau,Kinoshita2004TG}.
Under the tight transverse confinement applied to realize one-dimensional systems,
virtual transverse excitations induce effective multi-body interactions~\cite{PhysRevLett.100.210403}.
Although such corrections are often neglected in dilute gases,
they can play an important role in dynamics,
because they break the integrability of the ideal Lieb-Liniger model~\cite{tanaka:2022E,Mazets_2010,PhysRevLett.100.210403}.

In the context of ultracold quantum gases,
the scattering dynamics of clusters have been explored in few-body systems~\cite{PhysRevLett.96.163201,PhysRevLett.93.090404}.
For instance,
in one-dimensional three-particle (atom-dimer) collisions,
the breakdown of integrability due to energy-dependent interactions has been shown to induce cluster dissociation and recombination~\cite{PhysRevLett.96.163201}.
Furthermore,
in three-dimensional setups,
exact analyses of four-particle (dimer-dimer) scattering have elucidated the stability of molecular Bose-Einstein condensates~\cite{PhysRevLett.93.090404}.
While these studies have shed light on nontrivial aspects of few-particle cluster dynamics,
extending the theoretical analysis to clusters of arbitrary numbers of particles remains challenging.
This problem is particularly relevant in quasi-one-dimensional systems,
where transverse confinement induces effective multi-body interactions
and can lead to confinement-induced resonances~\cite{PhysRevLett.100.210403,PhysRevLett.104.153203,PhysRevLett.81.938}.

In this paper, we investigate scattering between unequal-size clusters in a quasi-one-dimensional Bose gas.
In particular, we aim to elucidate the consequences of the effective three-body attraction for cluster scattering.
To this end, we first analyze the integrable limit with only the two-body attraction and then treat the three-body attraction as a perturbation.
Throughout this paper, we focus on elastic two-cluster processes, excluding cluster breakup and rearrangement reactions.
Exploiting the L\"ushcer formula and the integrability of the unperturbed Lieb-Liniger Hamiltonian~\cite{Luscher1986VolumeDependence, LUSCHER1991531,PhysRevD.95.054508,Pozsgay_2011,PhysRevA.94.053620},
we evaluate the cluster-cluster scattering lengths.
This analysis applies to systems with total particle number as large as \(N \sim 50\),
well beyond the three- and four-body cases, and reveals the leading effects of integrability breaking.
Throughout this paper, we set \(\hbar = 1\).

\section{Preliminaries}
We work with one-dimensional Bose gases with contact interactions, 
whose Hamiltonian is given by
\begin{align}
  H = \int_0^L dx \Bigg[ &\frac{1}{2m} \frac{\partial}{\partial x} \phi^\dagger (x) \frac{\partial}{\partial x} \phi (x)
  + c \, \phi^{\dagger 2} (x)\phi^2 (x)
  \notag \\
  &+ u \, \phi^{\dagger 3} (x) \phi^3 (x) \Bigg]
  .
  \label{eq:Bose_gas}
\end{align}
Here,
\(m\) is the particle mass,
\(\phi\) is the bosonic field operator,
\(L\) is the system size,
and
\(c\) and \(u\) are the coupling constants of the two-body and three-body contact interactions, respectively.
Throughout this paper, we focus on the attractive regime \(c < 0\) and \(u < 0\).
It is convenient to parametrize the two-body coupling as
\begin{align}
  c = -\frac{1}{ma},
\end{align}
where \(a>0\) is the one-dimensional two-body scattering length,
which serves as a typical length scale for the range of the effective interaction between clusters.
The corresponding Schr\"odinger equation is 
\begin{align}
    \Bigg[
        & \sum_{j;\, 1\leq j \leq N} \frac{-1}{2m} \frac{\partial^2}{\partial x_j ^2} 
        + 2\, c \, \sum_{j,l;\, 1\leq j < l \leq N} \delta(x_j - x_l )
    \notag \\ &
        +\, 6\, u \, \sum_{j,l,n;\, 1\leq j < l < n \leq N} \delta(x_j - x_l ) \delta(x_l - x_n )
    \Bigg]
    \Psi(\bm{x})
    \notag \\
    &=
    E\,
    \Psi(\bm{x})
    ,
    \label{eq:seq_23}
\end{align}
where \(E\) is the energy eigenvalue.
The quasi-one-dimensional Bose gas considered here can be realized by transversely confining three-dimensional bosons with a two-body contact interaction in a harmonic trap~\cite{PhysRevLett.87.130402,PhysRevLett.81.938}.
In this realization, the one-dimensional coupling constants \(c\) and \(u\) are related to the three-dimensional two-body coupling constant \(g_{3\mathrm{D}}\) and the transverse confinement frequency \(\omega_\perp\) at leading order as~\cite{PhysRevLett.100.210403,Mazets_2010}
\begin{align}
    & 
    c = \frac{m\omega_\perp }{2\pi } g_{3\mathrm{D}} + \mathcal{O}(g_{3\mathrm{D}} ^2 )
    , 
    \label{eq:2-body_3d}
    \\
    &
    u = -\frac{\log \frac{4}{3}}{2\pi^2 } m^2 \omega_\perp g_{3\mathrm{D}} ^2 + \mathcal{O}(g_{3\mathrm{D}} ^3 )
    .
    \label{eq:3-body_3d}
\end{align}

In the absence of the three-body interaction, the Bose gas in Eq.~\eqref{eq:Bose_gas} reduces to the Lieb-Liniger model~\cite{lieb:1963,lieb:1963-2}.
In this model, 
the exact \(N\)-body wave function is obtained via the Bethe ansatz
as
\begin{align}
  & \Psi_{\mathrm{LL}} (\lambda_1 , \cdots, \lambda_N |x_1, \cdots, x_N) 
  \notag \\
  =&\, 
    \sum_{\sigma;\, \sigma \in S_N } \prod_{j,l;\, 1\leq j < l \leq N} 
    \left[ 1 + \frac{2mic \, \mathrm{sgn} (x_j - x_l )}{\lambda_{\sigma_j} - \lambda_{\sigma_l}} \right]
    \notag \\
    &\, \times
    \exp \left[i \sum_{n;\, 1\leq n \leq N} \lambda_{\sigma_n} x_n \right]
    ,
    \label{eq:wf_LL}
\end{align} 
where each \(\lambda_j\) is a so-called rapidity~\cite{lieb:1963,lieb:1963-2,KorepinBOOK}.
Under the periodic boundary condition, 
rapidities are quantized by the Bethe equations,
\begin{align}
  e^{i\lambda_j L} = - \prod_{l;\, 1\leq l \leq N} \frac{\lambda_j - \lambda_l + 2mic}{\lambda_j - \lambda_l - 2mic}
  ,
\end{align}
and,
in the attractive regime,
arrange themselves into a regular pattern in the complex plane~\cite{10.1063/1.1704156,Takahashi1999}.

The rapidities form groups that share a common real part and have equally spaced imaginary parts, and they are parameterized as
\begin{align}
  &
  \{\lambda_j\}_{j=1}^{N}
  = \bigcup_{\mu=1}^{N_{\mathrm{s}}} \{\lambda_{\mu,\nu}\}_{\nu=1}^{\alpha_\mu},
  \\
  &
  \lambda_{\mu,\nu}
  = \lambda_\mu^{\mathrm{s}}
   + i m c\, (\alpha_\mu + 1 - 2\nu)
   + i\delta_{\mu,\nu}^{\mathrm{s}}
  .
  \label{eq:string_hypothesis}
\end{align}
Here, \(N=\sum_{\mu=1}^{N_{\mathrm{s}}}\alpha_\mu\).
Each group \(\{\lambda_{\mu,\nu}\}_{\nu=1}^{\alpha_\mu}\) is called a string and constitutes a bound cluster of \(\alpha_\mu\) particles.
The common real part \(\lambda_\mu^{\mathrm{s}}\) is referred to as the string center and is determined by the Bethe-Gaudin-Takahashi equations~\cite{Takahashi1999,Calabrese_2007}:
\begin{align}
  &
  \alpha_\mu \lambda_\mu^{\mathrm{s}} L
  - 
  \sum_{\nu=1}^{N_{\mathrm{s}}}
  \Phi_{\mu,\nu}\!\left(\lambda_\mu^{\mathrm{s}}-\lambda_\nu^{\mathrm{s}}\right)
  = 2\pi I_\mu,
  \\
  &
    \begin{aligned}[b]
        \Phi_{\mu,\nu}&(\lambda)
        =
        -2\Bigg[
            (1-\delta_{\mu,\nu})\arctan\!\left(\frac{\lambda}{mc|\mu-\nu|}\right)
            \\
            & + \arctan\!\left(\frac{\lambda}{mc(\mu+\nu)}\right)
            \\
            & + 2\sum_{\alpha;\,1\le\alpha\le\min\{\mu,\nu\}-1}
            \arctan\!\left(\frac{\lambda}{mc(|\mu-\nu|+2\alpha)}\right)
  \Bigg]
    \end{aligned}
    ,
\end{align}
where \(I_\mu\) takes half-odd-integer (integer) values when the number of strings with length \(\alpha_\mu\) is even (odd).
From these rapidities, the total momentum and energy are given by
\begin{align}
  P &= \sum_{\mu=1}^{N_{\mathrm{s}}}\alpha_\mu \lambda_\mu^{\mathrm{s}},
  \\
  E &= \sum_{\mu=1}^{N_{\mathrm{s}}}
  \left[
    \frac{\alpha_\mu}{2m}(\lambda_\mu^{\mathrm{s}})^2
    - \frac{1}{3}m^2c^2\alpha_\mu(\alpha_\mu-1)
  \right].
\end{align}
The Bethe ansatz associated with the parameterization in Eq.~\eqref{eq:string_hypothesis}
is often referred to as the string ansatz.

So far, our discussion has been restricted to the Lieb-Liniger Bose gas.
In this case, the wave function can be determined exactly, which allows the scattering phase shift to be extracted at arbitrary momentum, as shown below in Eq.~\eqref{eq:ps_LL}.
However, this strategy is specific to integrable models and does not apply to generic systems.
Instead, one can use the L\"uscher formula, which relates the finite-volume energy spectrum to the scattering phase shifts at the corresponding relative momentum~\cite{Luscher1986VolumeDependence, LUSCHER1991531}.
For two distinguishable particles in one dimension, it takes the form of~\cite{PhysRevD.95.054508}
\begin{align}
  \cos \theta
  =
  \frac{\cos \left(\delta_+ + \delta_- + qL\right)}
       {\cos \left(\delta_+ - \delta_-\right)}
  .
  \label{eq:multi-channel_quantization}
\end{align}
Here, \(q\) is the relative momentum, \(L\) is the system size, and \(\delta_\pm(q)\) are the even- and odd-channel phase shifts.
In addition, \(\theta\) is defined by
\(\theta = -\alpha_1 QL/(\alpha_1 + \alpha_2)\),
where \(Q\) is the total momentum.
The equivalent choice \(\theta = \alpha_2 QL/(\alpha_1 + \alpha_2)\) leads to the same quantization condition for \(q\), 
since it leaves \(\cos\theta\) unchanged.

When the effective interaction between clusters is parity invariant,
the phase shifts admit the effective-range expansions at low relative momentum \(q \ll 1/a\)~\cite{PhysRev.76.38}.
In one dimension, they take the forms of~\cite{Vania}
\begin{align}
  q \tan \delta_+
  &= \frac{1}{a_+} + \frac{1}{2} r_+ q^2 + \mathcal{O}(a^4 q^4)
  ,
  \label{eq:ere_even}
  \\
  q \cot \delta_-
  &= \frac{-1}{a_-} + \frac{1}{2} r_- q^2 + \mathcal{O}(a^4 q^4)
  .
  \label{eq:ere_odd}
\end{align}
Here, \(a_\pm\) and \(r_\pm\) are the scattering lengths and effective
ranges in the even- and odd-parity channels, respectively.

\section{Two-cluster scattering}
In this section, we study two-cluster scattering in the one-dimensional Bose gas described by Eq.~\eqref{eq:Bose_gas}, focusing on the effects of the three-body attraction.
We consider the regime in which two clusters form and analyze their cluster-cluster scattering within the elastic approximation, neglecting cluster breakup and recombination.
Exploiting the L\"uscher formula together with the determinant formula for the three-body local correlation in the Lieb-Liniger Bose gas, 
we aim to evaluate the scattering lengths.
The scattering phase shifts entering the L\"uscher formula receive contributions from both the two-body and three-body interactions.
To isolate the effect of the three-body interaction on scattering, we first determine the phase shifts in the absence of the three-body interaction and then remove the corresponding two-body contribution from the L\"uscher formula.
We therefore begin with two-cluster scattering in the attractive Lieb-Liniger Bose gas.

\subsection{Attractive Lieb-Liniger Bose gas}
We begin with the scattering of two strings.
The exact wave function is given by Eq.~\eqref{eq:wf_LL}, 
with the rapidities split into two sets:
\begin{align}
  &
  \lambda_{1,\mu} = k + imc \, (\alpha_1 + 1 - 2 \mu ) + i \delta^{\mathrm{s}} _{1,\mu}
  ,
  \label{eq:string_1}
  \\
  &
  \lambda_{2,\mu} = p + imc \, (\alpha_2 + 1 - 2 \mu ) + i \delta^{\mathrm{s}} _{2,\mu}
  .
  \label{eq:string_2}
\end{align}
We denote the corresponding sets of rapidities by
\(\bm{\lambda}_1 = \{\lambda_{1,\mu}\}_{\mu=1}^{\alpha_1}\)
and
\(\bm{\lambda}_2 = \{\lambda_{2,\mu}\}_{\mu=1}^{\alpha_2}\).
Here, 
\(\alpha_1\) and \(\alpha_2\) are the string lengths and
\(k\) and \(p\) are the string centers of the two strings.
The string centers are related to each other via the following Bethe-Gaudin-Takahashi equation~\cite{Calabrese_2007}: 
\begin{align}
  &
  \alpha_1 k L = 2\pi I_1 + \Phi_{\alpha_1 ,\alpha_2 } (k-p)
  \label{eq:BGT_1}
  ,\\
  &
  \alpha_2 p L = 2\pi I_2 - \Phi_{\alpha_1 ,\alpha_2 } (k-p)
  \label{eq:BGT_2}
  ,
\end{align}
where \(I_1\) and \(I_2\) are the quantum numbers associated with the two string centers.
From these equations, we have
\begin{align}
  &
  Q = \frac{2\pi}{L} (I_1 + I_2 ) = \alpha_1 k + \alpha_2 p
  \label{eq:total_momentum}
  ,\\
  &
  q_{\mathrm{LL}} L = 2\pi \frac{\alpha_2 I_1 - \alpha_1 I_2 }{\alpha_1 + \alpha_2 } + \frac{\alpha_1 \alpha_2 }{\alpha_1 + \alpha_2 } \Phi_{\alpha_1 ,\alpha_2 } \left(\frac{\alpha_1 + \alpha_2 }{\alpha_1 \alpha_2 } q_{\mathrm{LL}} \right)
  ,
  \label{eq:BGT_rela}
\end{align}
where \(Q\) is the total momentum of the system
and \(q_{\mathrm{LL}}\) is the relative momentum between two strings, 
defined by
\begin{align}
  q_{\mathrm{LL}} \coloneqq \frac{\alpha_1 \alpha_2 }{\alpha_1 + \alpha_2 } (k-p)
  .
\end{align}

For scattering between two strings,
the asymptotic form of Eq.~\eqref{eq:wf_LL} at large string separation consists of a single plane-wave component, 
so that the reflection amplitude vanishes~\cite{PhysRev.168.1920}.
Consequently, the even- and odd-channel phase shifts agree modulo~\(\pi\), 
and are explicitly given by~\cite{10.1063/1.1704156,PhysRev.168.1920}
\begin{align}
    \delta_{\mathrm{LL},\pm}(q)\equiv
    \frac{\pi}{2}\delta_{\alpha_1,\alpha_2}
    -\Phi_{\alpha_1,\alpha_2}\!\left(\frac{\alpha_1+\alpha_2}{\alpha_1\alpha_2}q\right),
    \quad \mathrm{mod}\,\pi
    .
    \label{eq:ps_LL}
\end{align}

From this point onward,
we focus on the case where the string lengths are different.
In this case, according to Eqs.~\eqref{eq:ere_even} and \eqref{eq:ps_LL},
the inverse even-channel scattering length vanishes:
\begin{align}
  \frac{1}{a_{\mathrm{LL},+} } =& \lim_{q\to 0} q \tan \delta_{\mathrm{LL},+} (q) 
  = 0
  .
\end{align}
On the other hand, the even-channel effective range (\(r_{\mathrm{LL},+}\)) and odd-channel scattering length (\(a_{\mathrm{LL},-}\))
satisfy
\begin{align}
  r_{\mathrm{LL},+} = -2 \, a_{\mathrm{LL},-}
  ,
  \label{eq:er_even_and_sl_odd}
\end{align}
where \(r_{\mathrm{LL},+}\) and \(a_{\mathrm{LL},-}\) are defined by 
\begin{align}
  & r_{\mathrm{LL},+} = \lim_{q\to 0} \frac{\partial^2 }{\partial q^2 } q \tan \delta_{\mathrm{LL},+} (q)
  , \\
  & \frac{-1}{a_{\mathrm{LL},-}} = \lim_{q\to 0} q \cot \delta_{\mathrm{LL},-} (q)
  .
\end{align}
With these scattering parameters, Eq.~\eqref{eq:BGT_rela} reduces to
\begin{align}
  2\pi \frac{\alpha_2 I_1 - \alpha_1 I_2}{\alpha_1 + \alpha_2}
  =
  q_{\mathrm{LL}} (L + r_{\mathrm{LL},+} )
  \label{eq:BGT_rela_2}
\end{align}
in the low-energy regime,
where we used the effective-range expansion.
It is worth noting that the equal-string-length case is qualitatively different in that the inverse even-channel scattering length does not vanish.
This difference stems from whether the first term survives in Eq.~\eqref{eq:ps_LL}.

\subsection{Quasi-one-dimensional Bose gas}
Next, we turn to two-cluster scattering including the perturbative three-body interaction,
which constitutes the main part of this paper.
In what follows, we continue to assume unequal cluster sizes 
and work with the Schr\"odinger equation in Eq.~\eqref{eq:seq_23}.
The energy eigenvalue up to first order in the three-body attraction is given by
\begin{align}
    E 
    =&\, 
      \frac{\alpha_1 k^2}{2m} - \frac{m^2 c^2}{3} \alpha_1 (\alpha_1 ^2 - 1)
     + \frac{\alpha_2 p^2 }{2m} - \frac{m^2 c^2}{3} \alpha_2 (\alpha_2 ^2 - 1)
    \notag \\
    & + u \, C_3 \left(\bm{\lambda}_{1} \cup \bm{\lambda}_{2} \right)
    + \mathcal{O}(u ^2)
    .
    \label{eq:energy_entire_3}
\end{align}
Here, we introduced the local three-body correlation function
\begin{align}
    C_3 \left(\bm{\lambda}_{1} \cup \bm{\lambda}_{2} \right) 
    = \int_0^L dx\, \braket{\bm{\lambda}_{1} \cup \bm{\lambda}_{2}|\, \phi^{\dagger 3} (x)\phi^3 (x)\, |\bm{\lambda}_{1} \cup \bm{\lambda}_{2}}
    .
    \label{eq:three-body_contact_corr_definition}
\end{align}
In this expression, \(\ket{\bm{\lambda}_{1}\cup \bm{\lambda}_{2}}\) denotes the normalized Bethe eigenstate labeled by the rapidities in Eqs.~\eqref{eq:string_1} and \eqref{eq:string_2}.
Its coordinate-space representation is obtained from Eq.~\eqref{eq:wf_LL} up to normalization.

The energy in Eq.~\eqref{eq:energy_entire_3} contains the relative kinetic energy between the clusters, the binding energy of each cluster, the interaction energy between the clusters, and the total center-of-mass kinetic energy. 
By subtracting the binding and center-of-mass contributions, we obtain the relative-motion energy as
\begin{align}
    E_{\text{RM}} 
    =&\, E + \frac{1}{3} m^2 c^2 \alpha_1 (\alpha_1 ^2 - 1) + \frac{1}{3} m^2 c^2 \alpha_2 (\alpha_2 ^2 - 1)
    \notag \\
    &- E_{\text{3,bind}} - \frac{Q^2}{2m(\alpha_1 + \alpha_2)}
    \\
    =&\, \frac{\alpha_1 \alpha_2}{2m (\alpha_1 + \alpha_2)} (k-p)^2 + \delta E_3 
    + \mathcal{O}(m^2 u^2)
    \label{eq:energy_rel_3}
\end{align}
up to first order in the three-body coupling.
Here, \(Q\) denotes the total momentum, which is unaffected by the three-body perturbation. 
Therefore, the total momentum is still given by
\begin{align}
    Q = \alpha_1 k + \alpha_2 p
    ,
    \label{eq:momentum_tot_3}
\end{align}
with \(k\) and \(p\) being the centers of the two strings in Eqs.~\eqref{eq:string_1} and \eqref{eq:string_2}.
In addition, \(\delta E_3\) is the first-order energy shift after subtracting the three-body binding contribution, given explicitly by
\begin{align}
    \delta E_3
    = u\, \delta C_3 
    ,\quad 
    \delta C_3 
    \coloneqq C_3 (\bm{\lambda}_{1}\cup\bm{\lambda}_{2}) - C_{3,\text{const}}
    .
\end{align}
Here, \(C_{3,\text{const}}\) is the \(\mathcal{O}(L^0)\) term in the asymptotic form of \(C_3(\bm{\lambda}_{1}\cup\bm{\lambda}_{2})\) at large \(L\),
and \(E_{\text{3,bind}} = u\, C_{3,\text{const}}\) is the intra-cluster three-body contribution to the binding energies of the two clusters.
Accordingly, the relative momentum is found to be
\begin{align}
    q 
    =&\, \sqrt{2\mu_r E_{\text{RM}}}
    \notag \\
    =&\,
    \sqrt{
        \left[\frac{\alpha_1 \alpha_2 }{\alpha_1 +\alpha_2} (k-p)\right]^2
        + 2\mu_r \delta E_3
    }
    \,,
    \label{eq:momentum_rel_3}
\end{align}
where we selected the positive branch and employed the reduced mass of \(\mu_r \coloneqq \alpha_1 \alpha_2 m /(\alpha_1 + \alpha_2)\). 
The first term under the square root is the relative momentum squared in the absence of the three-body attraction, 
while the second term represents the energy shift induced by the three-body attraction.

We are now in a position to derive the even-channel scattering length in the presence of both two- and three-body interactions.
To this end, in the low-momentum regime \(a q \ll 1\), we first combine the multi-channel L\"uscher formula~\eqref{eq:multi-channel_quantization} with the effective-range expansions in Eqs.~\eqref{eq:ere_even} and~\eqref{eq:ere_odd}.
We then subtract Eq.~\eqref{eq:BGT_rela_2} from the resulting equation to isolate the effect of the three-body attraction,
relating the energy shift and the even-channel scattering length.

According to the L\"uscher formula~\eqref{eq:multi-channel_quantization} together with the effective-range expansions in Eqs.~\eqref{eq:ere_even} and~\eqref{eq:ere_odd},
we find 
\begin{align}
    \cos & \left(\frac{-\alpha_1 }{\alpha_1 + \alpha_2} Q L\right)
    \notag \\
    &=\,
    \frac{\cos \left(\frac{1}{a_+ q} - \left( a_- - \frac{1}{2} r_+ \right) q + qL\right)}{\cos \left(\frac{1}{a_+ q} + \left( a_- + \frac{1}{2} r_+ \right) q \right)}
    +
    \mathcal{O}(a^3 q^3)
    .
    \label{eq:ls_ere}
\end{align}
Here, \(a_\pm\) and \(r_\pm\) are the scattering lengths and effective ranges introduced in Eqs.~\eqref{eq:ere_even} and \eqref{eq:ere_odd},
while \(Q\) and \(q\) are the total and relative momenta defined in Eqs.~\eqref{eq:momentum_tot_3} and~\eqref{eq:momentum_rel_3}.

From this point onward, 
we restrict our attention to the regime in which the three-body coupling is sufficiently weak and the system size is much larger than the two-body scattering length, namely, \(|m u L/a| \ll 1\) and \(L/a \gg 1\). 
Then, it follows from Eq.~\eqref{eq:er_even_and_sl_odd} that
\begin{align}
    &
    \left( a_- - \frac{1}{2} r_+ \right) q = \mathcal{O} \left(\frac{a}{L}\right)
    ,
    \label{eq:scattering_parameters_q_negative}
    \\
    &
    \left( a_- + \frac{1}{2} r_+ \right) q = \mathcal{O} \left(\frac{m u a }{L}\right)
    ,
    \label{eq:scattering_parameters_q}
\end{align}
where we used \(q = \mathcal{O}(a/L)\) which holds for the low-momentum region. 
On the other hand, dimensional analysis implies
\begin{align}
    \frac{1}{a_+ q} = \mathcal{O}\!\left(\frac{m u L}{a}\right)
    .
    \label{eq:scattering_length_q}
\end{align}
We substitute Eqs.~\eqref{eq:scattering_parameters_q_negative}, \eqref{eq:scattering_parameters_q}, and~\eqref{eq:scattering_length_q} into Eq.~\eqref{eq:ls_ere}, with the aid of 
\(qL = \mathcal{O}((a/L)^0 )\),
to arrive at
\begin{align}
    \cos \left(\frac{-\alpha_1 }{\alpha_1 + \alpha_2} Q L\right)
    =&\,
    \cos \left(\frac{1}{a_+ q} - \left( a_- - \frac{1}{2} r_+ \right) q + qL\right)
    \notag \\
    &
    +
    \mathcal{O} \left( \frac{m u ^2 L^2}{a^2} \right)
    +
    \mathcal{O}(a^3 q^3)
    \label{eq:luscher_multi_3_2}
    .
\end{align}
This equation implies that the two cosine arguments are equal up to an overall sign and an integer multiple of \(2\pi\).
This ambiguity is resolved by selecting the branch consistent with the Bethe-Gaudin-Takahashi equation in the absence of the three-body attraction, yielding
\begin{align}
    2\pi \frac{ - \alpha_1 I_2 + \alpha_2 I_1 }{\alpha_1 + \alpha_2} 
    =
    \frac{1}{a_+ q} + q\, ( L + r_+ )
    \label{eq:luscher_ere}
\end{align}
up to \(\mathcal{O}\left(m u L/a, a q\right)\).

We are now ready to extract the even-channel scattering length.
In what follows, all expressions are understood as leading orders in \(m u L/a\) and \(a q\).
By subtracting Eq.~\eqref{eq:BGT_rela_2} from Eq.~\eqref{eq:luscher_ere},
we find
\begin{align}
    q_3 
    = -\frac{1}{q_{\mathrm{LL}} a_+ L }
    \label{eq:momentum_shift_3}
    .
\end{align}
Here, \(q_3 \coloneqq q - q_{\mathrm{LL}}\) denotes the shift of the relative momentum induced by the three-body attraction and satisfies
\begin{align}
    \frac{1}{2\mu_r} (q_{\mathrm{LL}} + q_3)^2 = E_{\text{RM}}
    ,
\end{align}
following from the definition of \(q_3\) and Eq.~\eqref{eq:momentum_rel_3}.
By substituting Eq.~\eqref{eq:momentum_shift_3} into Eq.~\eqref{eq:energy_rel_3},
we obtain
\begin{align}
    \delta E_3 = E_{\text{RM}} - \frac{1}{2\mu_r} q_{\mathrm{LL}} ^2 = \frac{1}{\mu_r} q_3 q_{\mathrm{LL}} = - \frac{1}{\mu_r a_+ L } 
    \label{eq:sl_even_energy_shift}
    ,
\end{align}
which may be viewed as a natural extension of Ref.~\cite{PhysRevA.97.061603}.
For numerical purposes, it is convenient to rewrite this relation in a dimensionless and coupling-free form,
\begin{align}
    \frac{m u a_+ }{a} = - \frac{1}{\mu_r a L \, \delta C_3 }
    ,
    \label{eq:sl_even_energy_shift_coupling_free}
\end{align}
whose right-hand side is evaluated in the next subsection.

\subsection{Numerical result}
We now turn to the numerical evaluation of the even-channel scattering length.
Equation~\eqref{eq:sl_even_energy_shift_coupling_free} shows that this evaluation requires the local three-body correlation function and its \(\mathcal{O}(L^0)\) part.
The former admits the following determinant representation~\cite{Pozsgay_2011,PhysRevA.94.053620}:
\begin{align}
    C_3 (\bm{\lambda}_{1}\cup\bm{\lambda}_{2})
    =&\,
    36\, L\,
    \sum_{B;\, B\subset \{1,\cdots,\alpha_1 + \alpha_2\} \, \land \, |B|=3}
    \frac{\operatorname{det} \mathscr{H}_B }{\operatorname{det} \mathscr{G}_B }
    \notag \\
    &\, \times
    \left[
        \prod_{j,l;\, 3 \geq j > l \geq 1}
        \frac{\lambda_{B,j} - \lambda_{B,l}}{(\lambda_{B,j} - \lambda_{B,l} )^2 + 4m^2 c^2 }
    \right]
    .
    \label{eq:local_three_body_correlation}
\end{align}
Here, we employed
\begin{align}
    \left[ \mathscr{H}_B \right]_{jl}
    \coloneqq 
    \begin{cases}
        \lambda_{B,j} ^{l-1} \quad \text{for } l=1,\cdots, 3 ,\\[2ex]
        \left[ \mathscr{G}_B \right]_{jl} \quad \text{for } l=4,\cdots, \alpha_1 + \alpha_2 ,
    \end{cases}
\end{align}
and the Gaudin matrix
\begin{align}
    \left[ \mathscr{G}_B \right]_{jl}
    \coloneqq 
    &\, 
    \delta_{jl}
    \left[
        L + \sum_{a;\, 1\leq a \leq \alpha_1 + \alpha_2 } \varphi (\lambda_{B,j} - \lambda_{B,a} )
    \right]
    \notag \\
    &
    - \varphi (\lambda_{B,j} - \lambda_{B,l} )
    ,
\end{align}
with \(\varphi\) being 
\begin{align}
    \varphi(\lambda) \coloneqq \frac{4mc}{\lambda^2 + 4m^2 c^2}
    .
    \label{eq:varphi_function}
\end{align}
In this notation, \(\lambda_{B}\) denotes the ordered set of rapidities obtained by reordering \(\bm{\lambda}_{1} \cup \bm{\lambda}_{2}\), with its first three entries given by \(\lambda_{B,j}\) for \(j=1,2,3\) and the remaining entries arranged in ascending order. 
It is worth noting that Eq.~\eqref{eq:local_three_body_correlation} apparently diverges because the rapidities within a perfect string are evenly spaced by \(2mc\)~\cite{Calabrese_2007}.
The divergence is resolved by taking into account the string deviations~\cite{KirillovKorepin1988_NormsBoundStates,Calabrese_2007}.
Since the string deviations decay exponentially with the system size \(L\), 
all finite-size corrections in negative powers of \(L\) are contained in the leading contribution.

The \(\mathcal{O}(L^0)\) part of the local three-body correlation function can be obtained straightforwardly
by evaluating the expectation value of the local three-body operator in the coordinate-space representation of
\(\ket{\bm{\lambda}_{1}\cup \bm{\lambda}_{2}}\).
Then, we arrive at 
\begin{align}
    C_{3,\text{const}} 
    =&\,
    4\, m^2 c^2
    \notag \\
    &\, \times
    \Bigg[
        \sum_{j=2}^{\alpha_1 -1} F_j ^{\alpha_1} F_{j+1} ^{\alpha_1}
        +
        \sum_{l=2}^{\alpha_2 -1} F_l ^{\alpha_2} F_{l+1} ^{\alpha_2}
    \Bigg] 
    +
    \mathcal{O}(L^{-1} )
    ,
    \label{eq:binding_3_body_contact}
\end{align}
where we employed \(F^{\mu} _{\nu} \coloneqq \nu (\mu - \nu)\).

\begin{figure}[t]
    \includegraphics[width=\columnwidth]{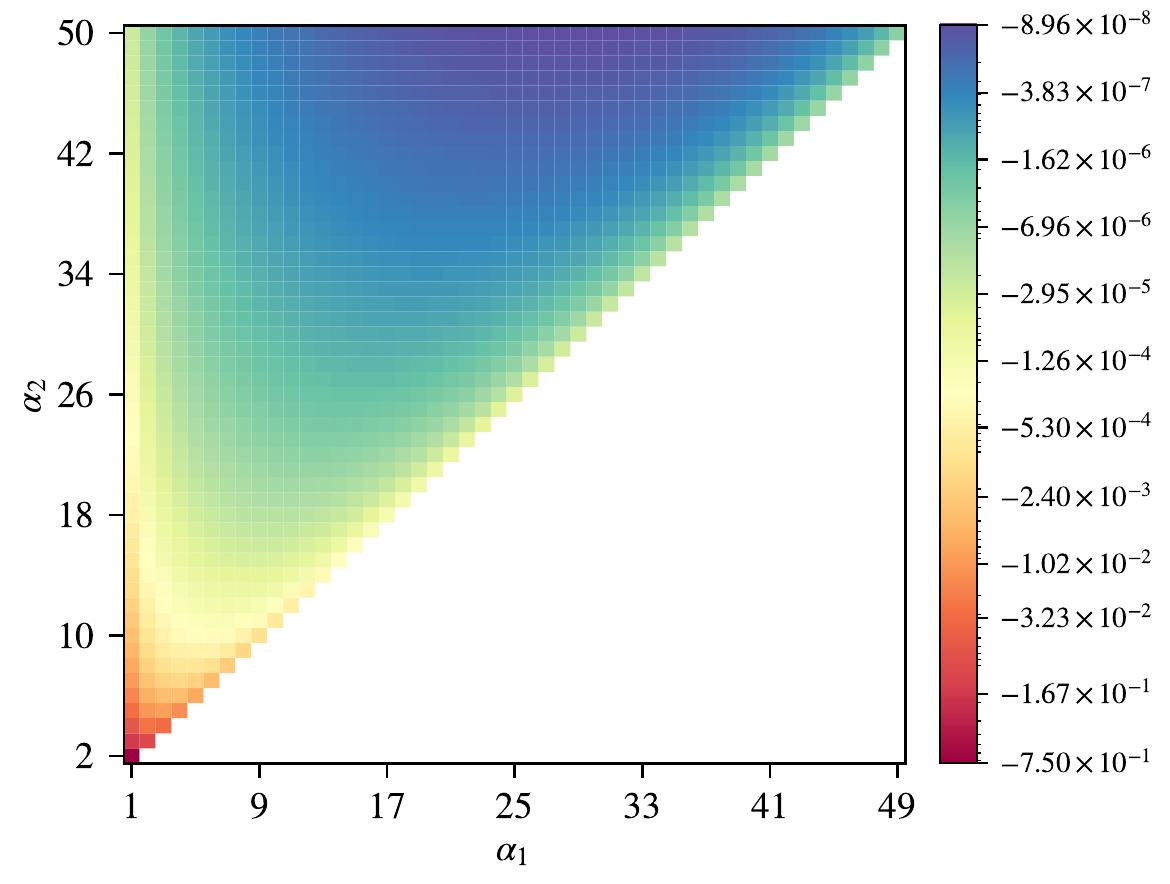}
    \caption{
        Heat map of the even-channel scattering length in the dimensionless and coupling-free form, \(m u a_+/a\), as a function of the cluster sizes.
        The sizes of the first and second clusters are denoted by \(\alpha_1\) and \(\alpha_2\), respectively, 
        while the value of \(m u a_+/a\) is represented by the color scale.
        Since the even-channel scattering length is invariant under exchanging the cluster sizes, 
        we show the heat map only for \(\alpha_2 > \alpha_1\).
        Throughout this figure, we set \(m = 1/2\) and \(c = -2\),
        where the scattering length for the original two-body attraction becomes \(a = 1\).
    }
    \label{fig:even_scattering_length_heatmap}
\end{figure}

We are now ready to numerically compute the even-channel scattering length according to Eq.~\eqref{eq:sl_even_energy_shift_coupling_free}.
First, we solve the Bethe-Gaudin-Takahashi equations in Eqs.~\eqref{eq:BGT_1} and \eqref{eq:BGT_2}
for the chosen quantum numbers and string lengths,
thereby determining the string centers.
We then construct the rapidities via Eqs.~\eqref{eq:string_1} and \eqref{eq:string_2}.
Next, we substitute the resulting rapidities into the determinant representation of the local three-body correlation function in Eq.~\eqref{eq:local_three_body_correlation}.
We then subtract \(C_{3,\text{const}}\) given by Eq.~\eqref{eq:binding_3_body_contact} to obtain \(\delta C_3\).
By substituting this \(\delta C_3\) into Eq.~\eqref{eq:sl_even_energy_shift_coupling_free}, we obtain the dimensionless and coupling-free even-channel scattering length, \(m u a_+/a\), 
which is presented as a heat map in Fig.~\ref{fig:even_scattering_length_heatmap}.
Representative numerical values underlying this heat map are listed in Tab.~\ref{tab:scatt_lengths_rep} in Appendix.
For the scattering between one particle and a two-body cluster, the resulting scattering length agrees with the previous result reported in Ref.~\cite{PhysRevA.97.061603}.

In the quasi-one-dimensional realization described by
Eqs.~\eqref{eq:2-body_3d} and \eqref{eq:3-body_3d},
the three-body interaction is always attractive.
Accordingly, Fig.~\ref{fig:even_scattering_length_heatmap} shows that
the even-channel scattering length is positive throughout,
indicating an attractive effective interaction between clusters.
This result has qualitatively different physical consequences for particle-cluster and cluster-cluster scatterings.
In the former case,
there is no energy continuum below the particle-cluster threshold,
so that a positive scattering length signals the formation of a bound state.
In the latter case,
the two-cluster threshold is embedded in an energy continuum,
so that a positive scattering length instead indicates the emergence of a resonance.
In both situations,
the associated binding energy 
is given by
\begin{align}
    E_{\text{bind}} = -\frac{1}{2\mu_r a_{+}^2},
    \label{eq:binding_energy_cluster_cluster}
\end{align}
which is of the same order as the decay width of resonance.

\section{Summary}
In this paper, we have investigated scattering between unequal-size clusters in a quasi-one-dimensional Bose gas.
When only the two-body contact attraction is present,
the even-channel scattering length diverges.
The situation changes in the quasi-one-dimensional setup considered here.
The weak three-body attraction induced by transverse confinement renders the even-channel scattering length finite and positive.
This result was obtained by exploiting the L\"uscher formula in Eq.~\eqref{eq:multi-channel_quantization}
together with the determinant representation of the local three-body correlation function in Eq.~\eqref{eq:local_three_body_correlation}.
The determinant representation allows us to go beyond the three- and four-body setting
up to the few-body regime of \(N \sim 50\).

The relationship between the even-channel scattering length and the energy shift is summarized in Eq.~\eqref{eq:sl_even_energy_shift},
and the resulting even-channel scattering length is presented in Fig.~\ref{fig:even_scattering_length_heatmap}.
Its positivity implies that the attractive effective interaction between clusters stems from the perturbative three-body attraction induced in the quasi-one-dimensional setup.
This result has different consequences for particle-cluster and cluster-cluster scatterings.
For scattering between a particle and a cluster,
a positive scattering length signals the formation of a bound state.
On the other hand, for scattering between unequal-size clusters,
the two-cluster threshold is embedded in an energy continuum,
so that a positive scattering length instead indicates the emergence of a resonance rather than a bound state.
In both cases,
the corresponding binding energy appears at second order in the three-body attraction,
and its explicit expression is given in Eq.~\eqref{eq:binding_energy_cluster_cluster}.

To conclude,
the perturbative three-body attraction induced in the quasi-one-dimensional setup yields a finite and positive even-channel scattering length,
thereby generating a bound state or a resonance in scattering between unequal-size clusters.
These results suggest that even a perturbatively weak quasi-one-dimensionality can govern the dynamical properties of the system through integrability breaking rather than integrable interactions.
A natural extension of the present analysis is to go beyond first order in the perturbative three-body attraction.
In particular,
the physical consequences derived from splitting and recombination processes mediated by the resonance are of interest.
At second order,
cluster recombination and breakup processes are expected to enter,
so that the elastic two-cluster description and the corresponding two-body L\"uscher formula are no longer sufficient.
Clarifying this regime will require a finite-volume quantization condition for coupled channels associated with different cluster partitions. 
The extended analysis may reveal how quasi-one-dimensionality drives the relaxation dynamics.

\begin{acknowledgments}
  The authors thank Shoki Sugimoto for valuable discussion.
  This work was supported by JSPS KAKENHI Grant No.~JP21K03384 and Matsuo Foundation.
\end{acknowledgments}

\appendix*
\section{Numerical data for Fig.~\ref{fig:even_scattering_length_heatmap}}
For reference, Tab.~\ref{tab:scatt_lengths_rep} lists representative numerical values underlying the heat map in Fig.~\ref{fig:even_scattering_length_heatmap}.

\begin{table*}[t]
   \begin{tabular}{rrrrrrrrrrrrrrrr}
     \midrule\midrule
      $\ \alpha_1$ & $\quad \alpha_2$ & $\qquad-m u a_+/a$ &$\ \ \qquad$& $\alpha_1$ & $\quad \alpha_2$ & $\qquad-m u a_+/a$ &$\ \ \qquad$& $\alpha_1$ & $\quad \alpha_2$ & $\qquad-m u a_+/a$ &$\ \ \qquad$& $\alpha_1$ & $\quad \alpha_2$ & $\qquad-m u a_+/a$ & $$ \\
     \midrule
      1 &  2 & 0.74997		&& 2 & 30 & 2.98281e-5	&& 8  &  9 & 0.00158002		&& 12 & 30 & 1.44406e-6 \\
      1 &  3 & 0.166663		&& 2 & 50 & 6.25500e-6	&& 8  & 10 & 0.00045819		&& 12 & 50 & 2.28095e-7 \\
      1 &  4 & 0.0624992	&& 3 &  4 & 0.0323268	&& 8  & 11 & 0.000212024	&& 20 & 21 & 8.74629e-5 \\
      1 &  5 & 0.0299997	&& 3 &  5 & 0.0085713	&& 8  & 20 & 1.09524e-5		&& 20 & 22 & 2.60023e-5 \\
      1 &  6 & 0.0166665	&& 3 &  6 & 0.00364124	&& 8  & 30 & 2.48497e-6		&& 20 & 23 & 1.24920e-5 \\
      1 &  7 & 0.010204		&& 3 & 10 & 0.000503302	&& 8  & 50 & 4.54099e-7		&& 20 & 30 & 1.24019e-6 \\
      1 & 10 & 0.00333332	&& 3 & 20 & 4.96836e-5	&& 10 & 11 & 0.000783726	&& 20 & 50 & 1.12360e-7 \\
      1 & 12 & 0.00189393	&& 3 & 30 & 1.37993e-5	&& 10 & 12 & 0.00022962		&& 30 & 31 & 2.41406e-5 \\
      1 & 20 & 0.000394736	&& 3 & 50 & 2.84294e-6	&& 10 & 13 & 0.000107614	&& 30 & 32 & 7.17744e-6 \\
      1 & 30 & 0.000114942	&& 4 &  5 & 0.0135306	&& 10 & 20 & 9.07795e-6		&& 30 & 33 & 3.47032e-6 \\
      1 & 50 & 2.44898e-5	&& 4 &  6 & 0.00371282	&& 10 & 30 & 1.80173e-6		&& 30 & 50 & 9.01267e-8 \\
      2 &  3 & 0.10714		&& 4 &  7 & 0.00162477	&& 10 & 50 & 3.08194e-7		&& 40 & 41 & 9.68258e-6 \\
      2 &  4 & 0.0267852	&& 4 & 10 & 0.000348187	&& 12 & 13 & 0.000440979	&& 40 & 42 & 2.87232e-6 \\
      2 &  5 & 0.0108694	&& 4 & 20 & 3.00527e-5	&& 12 & 14 & 0.000129995	&& 40 & 43 & 1.39156e-6 \\
      2 & 10 & 0.000956628	&& 4 & 30 & 8.10310e-6	&& 12 & 15 & 6.14463e-5		&& 40 & 50 & 1.51473e-7 \\
      2 & 20 & 0.00010469	&& 4 & 50 & 1.63699e-6	&& 12 & 20 & 8.77479e-6		&& 49 & 50 & 5.08423e-6 \\
     \midrule\midrule
   \end{tabular}
   \caption{Representative dimensionless and coupling-free even-channel scattering lengths, \(-m u a_+/a\), for cluster sizes $1 \leq \alpha_1 < \alpha_2 \leq 50$.
  The corresponding heat map is shown in Fig.~\ref{fig:even_scattering_length_heatmap}.
  }
\label{tab:scatt_lengths_rep}
\end{table*}

\bibliography{reference}

\end{document}